\documentclass[12pt]{article}

\usepackage[a4paper,margin=1.15in]{geometry}
\usepackage{amsmath,amssymb,amsthm}
\usepackage{graphicx}
\usepackage{booktabs}
\usepackage{hyperref}
\usepackage{caption}
\usepackage{subcaption}
\usepackage{enumitem}
\usepackage[T1]{fontenc}
\usepackage[utf8]{inputenc}

\usepackage[numbers,sort&compress]{natbib}

\newtheorem{theorem}{Theorem}

\newcommand{\MPC}{\mathrm{MPC}}

\title{Pick-to-Learn Calibration of an MPC Policy for\\ an Origin-to-Destination Flight Problem}

\author{Marco C. Campi and Simone Garatti % <-this % stops a space
\thanks{M.C. Campi is with the Department of Information Engineering, University of Brescia, via Branze 38, 25123 Brescia, Italy. S. Garatti is with the Dipartimento di Elettronica, Informazione e Bioingegneria, Politecnico di Milano, piazza L. da Vinci 32, 20133 Milano, Italy. Email: {\tt marco.campi@unibs.it, simone.garatti@polimi.it}.}%
}

\date{}

\begin{document}
\maketitle

\begin{abstract}
This paper illustrates the Pick-to-Learn methodology applied to the calibration of a Model Predictive Control policy. While developed around a specific example, the presentation is meant to highlight a methodology of broad applicability. The example concerns an aircraft traveling from an origin point to a destination point in the presence of uncertain crosswinds and a low-connectivity zone that should be avoided. The MPC policy is parameterized by two hyperparameters, which are selected from data by the P2L procedure. Starting from a dataset of 400 wind realizations, also called scenarios, P2L identifies a final compression set containing only two informative scenarios. The resulting MPC policy avoids the low-connectivity zone on all available scenarios and, according to the P2L theory, satisfies a probabilistic risk bound of 4.8\% at confidence level $1-10^{-5}$, where the risk is the probability of entering the low-connectivity zone in a future flight under a new wind realization not included in the sample.
\end{abstract}

\section{Introduction}

Model Predictive Control (MPC) is a flexible methodology for feedback control design whose practical performance often depends on the choice of tuning parameters. These parameters may encode trade-offs between competing objectives, such as performance and constraint satisfaction. In applications affected by uncertainty, selecting such parameters is especially delicate because a controller that performs well on a nominal model may fail under future uncertainty realizations.

This paper illustrates the application of the recently introduced Pick-to-Learn (P2L) methodology to the calibration of an MPC policy. P2L is a technique developed within the framework of the scenario approach, \cite{CampiGaratti2018,CampiCareGaratti2021,CGL2023}. It stemmed from an idea originally introduced in \cite{CGR2015} and was subsequently developed into a full-fledged methodology in \cite{P2L2023,P2L2025,P2L2025ArXiv}. The policy is learned from a finite dataset of uncertainty realizations and is accompanied by a probabilistic certificate on future constraint satisfaction. The methodology is demonstrated on an origin-to-destination flight problem in which an aircraft must reach a destination while avoiding a low-connectivity zone in the presence of uncertain crosswinds. While this example is intentionally a toy problem, it serves the purpose of clearly illustrating the methodology and the theoretical certification results provided by P2L. This same framework can be ported to substantially more complex MPC applications. 

Previous work on the use of P2L for MPC can be found in \cite{Zuliani2025}, which we credit as the pioneering contribution in this direction. The same authors had also earlier opened a line of research for the calibration of MPC parameters through alternative techniques, such as backpropagation and gradient estimation in \cite{Zuliani2025BPMPC,pmlr-v331-zuliani26a,zulianietal2026-ArXiv}, based on the closed-loop trajectory generated by the parametrized MPC policy. While in \cite{Zuliani2025} P2L was seen as an instrument to evaluate the so-called ``complexity'' of the policy, which measures the level of data compression in policy construction, we introduce P2L here primarily as \emph{a means} to achieve compression: P2L is seen as an enabler rather than simply as a tool for computing complexity. Further, the goal of this paper is to bring to light the use of state-of-the-art results, tighter than those used in \cite{Zuliani2025}, for the certification of constraint satisfaction. Finally, building on the so-called P2L$^+$ methodology introduced in \cite{P2L2025ArXiv}, this paper develops a post-design certification step that provides a rigorous basis for the formal verification of essentially any property of interest.

In the flight problem, the MPC controller is specified by two hyperparameters, denoted by $\rho$ and $\nu$. For every fixed pair $(\rho,\nu)$, the repeated solution of a finite-horizon optimal control problem (FHOCP) defines a feedback policy, denoted by
\[
\pi = \MPC(\rho,\nu).
\]
P2L is then used to select $(\rho,\nu)$ from data. The desired property is avoidance of the low-connectivity zone under the wind realizations in the dataset. The final output is not only a calibrated MPC policy, but also a small compression set of scenarios that supports the probabilistic generalization guarantee.

\subsection{P2L for MPC}
\label{section P2L for MPC}

Let
\[
D=(\delta_1,\ldots,\delta_N)
\]
be a dataset of independent uncertainty realizations. The elements of $D$ are also called scenarios. In the flight example developed in Section \ref{sec:flight_example}, each scenario $\delta_i$ is one wind realization.

\begin{figure}[ht]
	%\vspace{1in}
	\centering
\includegraphics[width=13cm]{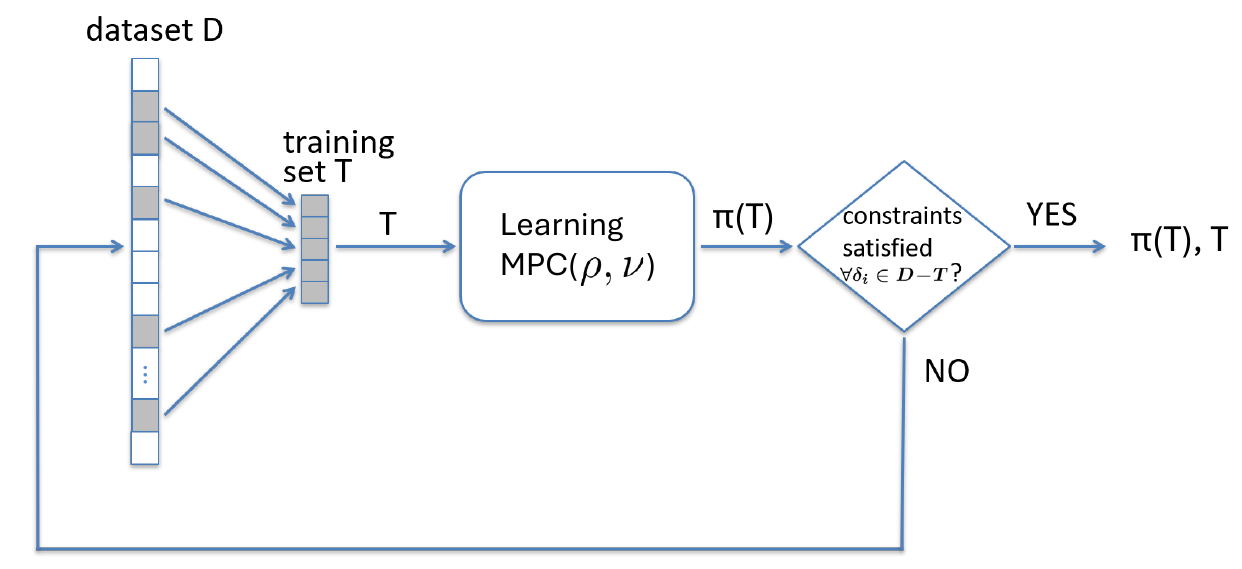}
\caption{P2L loop specialized to MPC calibration. The current training set $T$ is sent to the learning MPC block, which selects $(\rho,\nu)$ and returns the policy $\pi(T)=\MPC(\rho,\nu)$. The policy is checked on the remaining scenarios in $D\setminus T$. If a violation is found, one scenario is picked and added to $T$; otherwise, the algorithm comes to termination.}
\label{fig:p2l_loop}
\end{figure}

P2L incrementally constructs a training set $T\subseteq D$. Given the current training set $T$, a learning block selects the MPC tuning parameters and hence returns the policy
\[
\pi(T) = \MPC(\rho,\nu).
\]
The policy is then checked on the scenarios in $D\setminus T$. If the desired property is satisfied for all those scenarios, the procedure terminates and returns the pair $(\pi(T),T)$. Otherwise, one violating scenario is picked, added to $T$, and the learning step is repeated. The P2L procedure is illustrated in Figure \ref{fig:p2l_loop}. 

The P2L framework leaves two ingredients to the designer. The first is the learning block, namely the procedure that produces a policy from the current training set. As more precisely specified in the next section, in this paper the learning block tunes the MPC hyperparameters $(\rho,\nu)$ so as to minimize the distance traveled while avoiding the low-connectivity zone. The second is the rule used to choose which scenario in $D\setminus T$ should be added to $T$ when the current policy violates the desired property. In the flight example, the picking rule selects the wind realization producing the largest violation of the low-connectivity constraint.

The key point is that P2L returns both a policy $\pi(T)$ and a final training set $T$. This set is interpreted as a compression of the original dataset: it contains the scenarios that were actually used by the algorithm to construct the policy. By utilizing the theory of learning under sample compression of \cite{CGL2023}, P2L turns the size of $T$ into a probabilistic certificate.

Let $\varphi(\pi,\delta)\in\{0,1\}$ denote the property to be certified, where $\varphi(\pi,\delta)=1$ means that policy $\pi$ satisfies the property under the uncertainty realization $\delta$, and $\varphi(\pi,\delta)=0$ means that the property is violated. The risk of a policy is
\[
R(\pi)=\mathbb P_{\delta}\{\delta:\varphi(\pi,\delta)=0\},
\]
namely the probability that the property is violated for a new, previously unseen realization of the uncertainty.

The following result is the P2L generalization theorem used in this paper and borrowed from \cite{P2L2025ArXiv}.

\begin{theorem}[P2L generalization bound]\label{thm:p2l}
Let $(\pi(T),T)$ be the output returned by P2L when applied to a dataset $D$ of $N$ independent uncertainty realizations. For any confidence parameter $\beta\in(0,1)$, with confidence at least $1-\beta$ with respect to the random extraction of the dataset, the risk of the returned policy satisfies
\[
R(\pi(T))\leq \varepsilon(|T|,\beta,N),
\]
where, for $k=0,1,\ldots,N-1$, $\varepsilon(k,\beta,N)$ is the unique solution in $[k/N,1]$ of
\[
\frac{\beta}{N}\sum_{m=k}^{N-1}
\frac{\binom{m}{k}}{\binom{N}{k}}
(1-\varepsilon)^{-(N-m)}=1,
\]
and $\varepsilon(N,\beta,N)=1$.
\end{theorem}

For any given value of $\beta$ and $N$, function $\varepsilon(k,\beta,N)$ is increasing in $k$. Thus, the smaller the final training set $T$, the tighter the resulting certificate. In the flight example below, P2L is applied to $N=400$ wind realizations and terminates with $|T|=2$. Theorem~\ref{thm:p2l} is then used with $\beta=10^{-5}$ to obtain the bound
\[
R(\pi(T))\leq 4.8\%.
\]
Here, $R(\pi)$ is the probability that, under a fresh wind realization, the aircraft trajectory generated by the learned MPC policy enters the low-connectivity zone. Later in the paper, it is shown that the framework can also certify additional properties beyond constraint satisfaction through the extended methodology P2L$^+$. This include the satisfaction of more restrictive state constraints and an evaluation of the distance traveled by the aircraft. 

\section{Origin-to-Destination Flight Example} \label{sec:flight_example}

\subsection{Problem Description}

Consider an aircraft that must travel from the origin
\[
O=(0,0)
\]
to the destination
\[
D=(0,20).
\]
Throughout, we refer to the first coordinate as the horizontal coordinate and the second as the vertical coordinate. The nominal route is the vertical segment connecting these two points.

The flight is affected by crosswind disturbances acting within a wind zone that extends from vertical coordinate 1 to vertical coordinate 15. The wind is uncertain and varies from one flight to another. In the data-driven formulation adopted here, each element of the dataset is a realization of this wind field.

\begin{figure}[ht]
	%\vspace{1in}
	\centering
	\includegraphics[width=10cm]{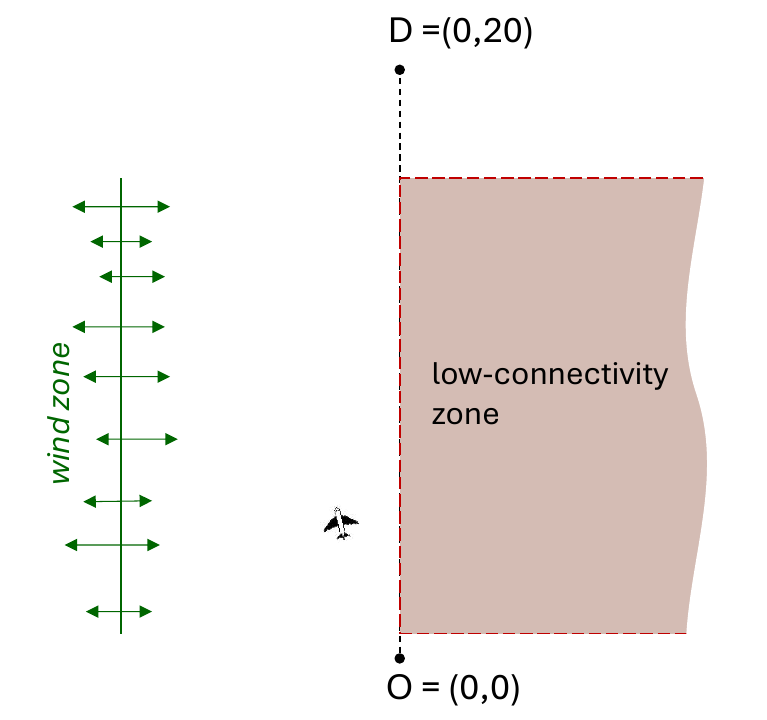}
\caption{Schematic representation of the origin-to-destination flight problem. The aircraft moves from $O=(0,0)$ to $D=(0,20)$ under uncertain crosswind disturbances. The low-connectivity zone is to be avoided.}
\label{fig:geometry}
\end{figure}

A low-connectivity zone is located close to the nominal route. If the aircraft enters this region, communication with ground infrastructure may deteriorate or be lost. The operational requirement is therefore that the aircraft should avoid this region, possibly at the price of traveling a slightly longer path. The setting is illustrated in Figure \ref{fig:geometry}. 

\subsection{Finite-Horizon Problem and MPC Policy}

The complete trip is discretized into 200 sampling intervals. At each sampling instant MPC solves a finite-horizon optimal control problem (FHOCP) over a horizon of 20 steps. Only the first control input is applied; at the next time instant the horizon is shifted forward and the optimization is solved again.

The aircraft is modeled as a point mass moving in a two-dimensional plane. We start from a continuous-time model that will be then discretized. The continuous-time state is
\[
x=(p_1,p_2,v_1,v_2),
\]
where $p_1$ and $p_2$ are ground-relative positions along the two orthogonal horizontal and vertical axes, and $v_1$ and $v_2$ are the corresponding air-relative velocity components. The wind is represented by a scalar disturbance $w$ acting along the first coordinate only. The continuous-time dynamics are
\begin{align*}
	\dot p_1 &= v_1+w, &
	\dot p_2 &= v_2,\\
	\dot v_1 &= \tau_1-0.5v_1, &
	\dot v_2 &= \tau_2-0.5v_2,
\end{align*}
where $\tau=(\tau_1,\tau_2)\in\mathbb R^2$ is the thrust command and the terms $-0.5 v_i$ are viscous drags. The thrust command is constrained by
\refstepcounter{footnote}
\[
\|\tau\|\leq 2.\footnotemark[\value{footnote}]
\]
\footnotetext[\value{footnote}]{While the linear model considered here may come across as overly simple, more realistic nonlinear models come down to a similar formulation after applying a feedback linearization, see e.g. \cite{Deori}.}
The model is discretized using a sample-and-hold approach with sampling time $T_s=0.1$. Thus, for $t\in[kT_s,(k+1)T_s)$, $\tau(t)=\tau_k$ and, in closed-loop simulations with wind, $w(t)=w_k$. With
\[
x_k=(p_{1,k},p_{2,k},v_{1,k},v_{2,k}),
\]
the resulting discrete-time model is
\begin{equation} 
x_{k+1}=Ax_k+B\tau_k+Ew_k,
\label{discrite-time model with wind}
\end{equation}
where 
\[
A=\begin{bmatrix}
	1.0000 & 0.0000 & 0.0975 & 0.0000\\
	0.0000 & 1.0000 & 0.0000 & 0.0975\\
	0.0000 & 0.0000 & 0.9512 & 0.0000\\
	0.0000 & 0.0000 & 0.0000 & 0.9512
\end{bmatrix},\qquad
B=\begin{bmatrix}
	0.0049 & 0.0000\\
	0.0000 & 0.0049\\
	0.0975 & 0.0000\\
	0.0000 & 0.0975
\end{bmatrix},
\]
and
\[
E=\begin{bmatrix}
	0.1000\\
	0.0000\\
	0.0000\\
	0.0000
\end{bmatrix}.
\]
Although the wind model is not used in the construction of the MPC policy, which is based only on wind realizations, we specify it here for completeness and reproducibility. The wind disturbance is denoted by $w_k(p_2)$. It depends on the discrete time $k$ and on the second position coordinate $p_2 \in [0,20]$, while it is assumed to be constant along the $p_1$ coordinate. Along the $p_2$ direction, the wind field is represented by a stochastic piecewise-linear function on a uniform grid with spacing $0.5$. Let
\[
p_{2,j}=\frac{j}{2}, \qquad j=0,\ldots,40,
\]
denote the grid points. At time $k=0$, the values $w_0(p_{2,j})$ are generated as a stationary realization of the autoregressive process
\[
w_0(p_{2,j+1}) = 0.95\, w_0(p_{2,j}) + \eta_j,
\qquad j=0,\ldots,39,
\]
where $\eta_j \sim \mathcal N(0,4\cdot 10^{-4})$ are independent Gaussian random variables. For $k>0$, the wind field evolves according to
\[
w_{k+1}(p_{2,j})
=
0.95\, w_k(p_{2,j})
+
\sum_{i=0}^{40} 0.9^{|p_{2,j}-p_{2,i}|}\eta_{k,i},
\qquad j=0,\ldots,40,
\]
where $\eta_{k,i} \sim \mathcal N(0,4\cdot 10^{-4})$. All Gaussian random variables appearing in the model are mutually independent. Values of $w_k(p_2)$ at points not belonging to the grid are obtained by linear interpolation. 

As previously described, the MPC policy is implemented over a total time horizon of 200 instants and relies on a finite-horizon optimal control problem (FHOCP) over a shorter prediction horizon of 20 steps, which is repeatedly solved at each sampling time according to a receding-horizon strategy. Moving to the description of the FHOCP, a wind-free model is used that is obtained from \eqref{discrite-time model with wind} by dropping the term $Ew_k$. For a given value of $(\rho,\nu)$, the FHOCP contains three terms in the objective function. The first term attracts the aircraft toward the destination. The second term attracts the horizontal position toward a virtual vertical line located at horizontal coordinate $\nu$. The third term penalizes the control effort. The parameter $\rho$ is the weight assigned to the virtual-line attraction term.

For a current state $x$, the FHOCP is written as
\begin{align}
\label{eq:fhocp_constraint} 
\min_{\tilde{\tau}_0,\ldots,\tilde{\tau}_{19}} \quad &
\sum_{j=1}^{20}\left(\|\tilde{p}_j-D\|^2+\rho(\tilde{p}_{1,j}-\nu)^2\right)
+\sum_{j=0}^{19}0.01\| \tilde{\tau}_j\|^2 \\
\text{subject to}\quad & \tilde{x}_{j+1}=A \tilde{x}_j+B \tilde{\tau}_j, \ \tilde{x}_0 = x, \qquad j=0,\ldots,19, \nonumber \\
& \nonumber \|\tilde{\tau}_j\|\leq 2,\qquad j=0,\ldots,19 
\end{align}
(we used the \textasciitilde~notation to distinguish the variables as predicted in FHOCP from the actual system variables). As already anticipated, the wind is not appearing in the FHOCP. It enters only at the level of the P2L calibration loop, where different wind scenarios are used to evaluate the closed-loop trajectories generated by candidate values of $(\rho,\nu)$. 

Given $(\rho,\nu)$, FHOCP returns an optimal input sequence, of which the first input $\tilde{\tau}_0$ is selected. Setting $\tau = \tilde{\tau}_0$ defines the feedback law
\[
\tau=\pi(x),
\]
and the MPC control scheme is achieved by using $\tau_k = \pi(x_k)$ in \eqref{discrite-time model with wind}.

In what follows the feedback policy will be also denoted by $\MPC(\rho,\nu)$ to emphasize its dependence on parameters $\rho,\nu$. The MPC design problem considered here is therefore reduced to the selection of these two hyperparameters $\rho$ and $\nu$, a step that is implemented through the P2L procedure as explained in Section \ref{section P2L for MPC} and precisely specified for the flight problem in the next Section \ref{P2L calibration for flight problem}.

\section{P2L Calibration of the MPC Hyperparameters}
\label{P2L calibration for flight problem}

The available dataset contains
\[
N=400
\]
wind field realizations. Initially, the training set $T$ is empty. The P2L procedure progressively augments $T$ by selecting wind scenarios that are informative for the design of the MPC policy.

For a given training set $T$, the learning block in Figure \ref{fig:p2l_loop} selects $(\rho,\nu)$ by solving
\begin{align}
\min_{\rho,\nu} \quad &
\frac{1}{|T|}\sum_{\delta_i\in T}
L(\MPC(\rho,\nu),\delta_i) \label{eq:p2l_learning_cost}\\
\text{subject to}\quad &
\varphi(\MPC(\rho,\nu),\delta_i)=1,
\qquad \forall\delta_i\in T. \label{eq:p2l_learning_constraint} \nonumber 
\end{align}
Here $L(\MPC(\rho,\nu),\delta_i)$ denotes the traveled distance of the closed-loop trajectory generated by the policy $\MPC(\rho,\nu)$ under wind scenario $\delta_i$. The constraint $\varphi(\MPC(\rho,\nu),\delta_i)=1$ requires that the corresponding trajectory does not enter the low-connectivity zone.

In the present example, the optimization problem in \eqref{eq:p2l_learning_cost} involves only two variables, $\rho$ and $\nu$, and is therefore easy to solve, even by a grid-based procedure. In this numerical example, we solved it using a standard Bayesian optimization tool available in MATLAB. More generally, the outer optimization associated with the learning block can be a challenging step that deserves specific attention. Importantly, the risk evaluation derived in this paper maintains its validity even when only an approximated solution is available.

After the learning step, the resulting policy is tested on all wind scenarios in $D\setminus T$. If no trajectory enters the low-connectivity zone, the procedure stops. If at least one trajectory violates the constraint, the next scenario is picked as the wind realization that produces the largest violation.

In this example, the violation magnitude is measured by the spatial area of the trajectory lying inside the low-connectivity region. The selected realization is inserted into $T$ and the learning step is repeated. This selection criterion is one possible choice. It is not imposed by the P2L theory. It is a user-defined rule intended to learn a good controller quickly by selecting an informative violating realization. 

\subsection{A Note on Robustification and Disturbance Handling}
We close this section with a broader discussion on how disturbances are typically handled in MPC, and contrast this with the role they play in the P2L-based procedure studied here.

Traditional robust or stochastic MPC design methods account for disturbances, when present, within the FHOCP, either explicitly through sampled disturbance scenarios, chance constraints, or min--max formulations, or indirectly through robustification mechanisms such as constraint tightening and tube-based constructions,  \cite{BemporadBorrelliMorari2003,LangsonChryssochoosRakovicMayne2004,MayneSeronRakovic2005,CannonChengKouvaritakisRakovic2012,PrandiniGarattiLygeros2012,SchildbachFagianoFreiMorari2014,FarinaGiulioniScattolini2016,CampiGarattiPrandini2018,DeoriGarattiPrandini2020,Rakovic2023ImplicitRigidTube,WangZhangGrosRakovic2025}. This is important because, if disturbances are ignored in the FHOCP, the resulting control law may be significantly affected by them once implemented in closed-loop. For instance, a disturbance may push the state toward a forbidden region. If the FHOCP is formulated without accounting for this effect, then a current state lying outside the forbidden region may nonetheless lead, shortly afterwards, to a constraint violation.

The P2L-based MPC procedure introduced here addresses this issue from a different perspective. The FHOCP itself does not include the disturbances. However, at the calibration level, P2L selects the MPC parameters $(\rho,\nu)$ so as to enforce constraint satisfaction on the closed-loop trajectories generated under the sampled disturbance realizations. In this sense, P2L provides an external, global supervision of the MPC policy: the FHOCP retains the basic receding-horizon structure of MPC and its implementation is further simplified by the fact that disturbances are not included at this level, while the effect of disturbances is accounted for by the P2L calibration performed at the level of complete closed-loop trajectories.

At the same time, it should be noted that nothing in the proposed framework prevents one from incorporating disturbances or other robustification mechanisms directly in the FHOCP. The point is rather that, within the present P2L-based approach, this is not strictly necessary in order to obtain a global probabilistic certificate for the resulting closed-loop behavior. 

\section{Numerical Results}

The procedure starts with
\[
\rho=0,
\qquad
\nu=-5.
\]
Since $\rho=0$, the virtual-line term is inactive. Without wind, the aircraft would fly along the straight path from the origin to the destination. Under wind, however, the trajectories are displaced, and some enter the low-connectivity zone.

\begin{figure}[ht]
	%\vspace{1in}
	\centering
	\includegraphics[width=14cm]{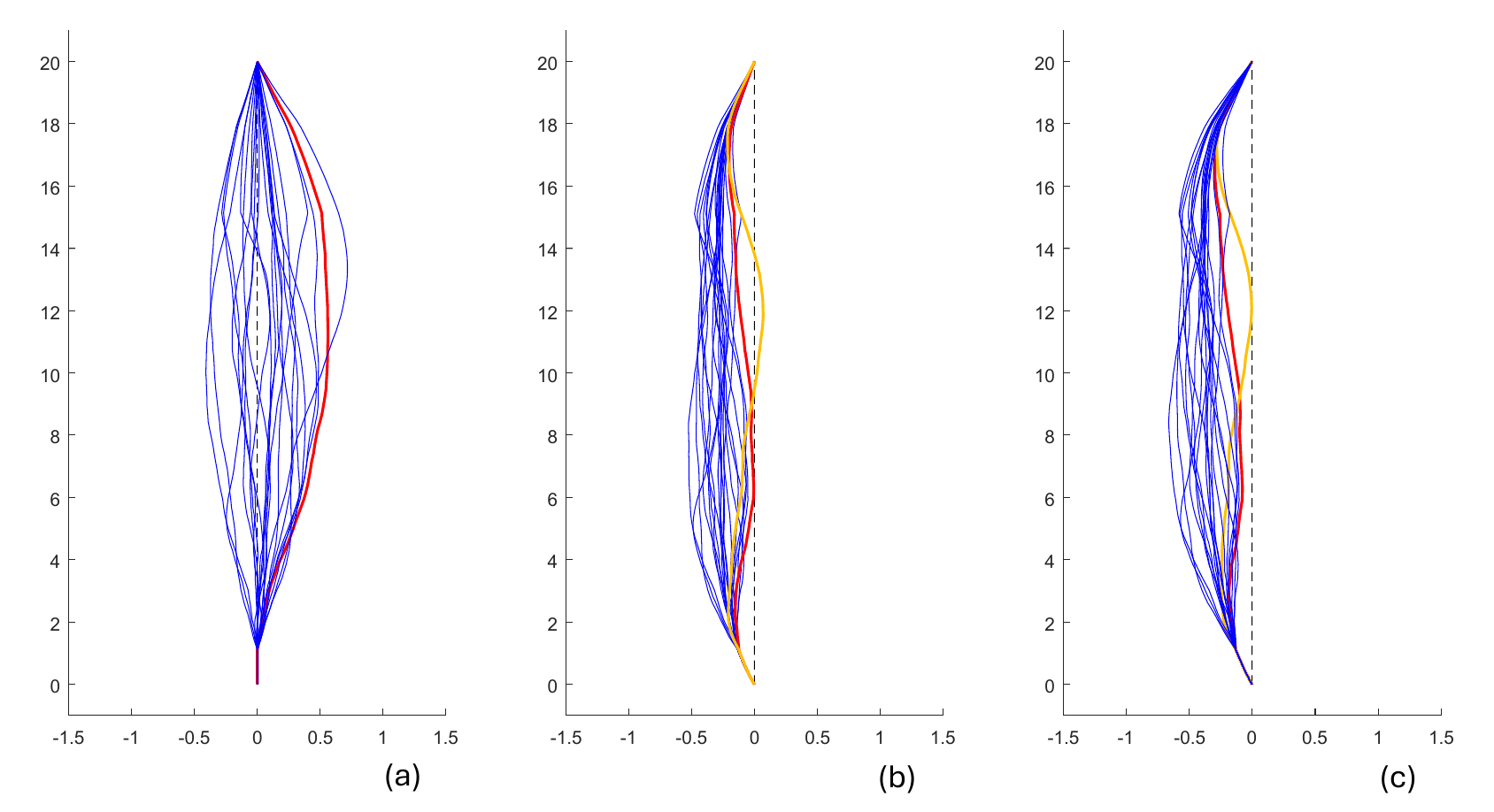}
\caption{Trajectories obtained at the three steps of P2L. Panel (a): trajectories with $\rho=0$ and $\nu=-5$. The red trajectory is the realization with the largest violation and is inserted into the training set. Panel (b): Trajectories after the first learning step. One remaining wind realization still violates the constraint and is added to the training set. Panel (c): Trajectories with the final policy obtained after two P2L picks. No trajectory violates the low-connectivity constraint.}
\label{fig:first_pick}
\end{figure}

Panel (a) in Figure~\ref{fig:first_pick} shows a representative subset of 20 trajectories among the 400 available. The red trajectory is the worst one according to the violation criterion. It is selected by P2L and inserted into the training set.

At this stage $|T|=1$. The learning block optimizes $\rho$ and $\nu$ using only this wind realization. The optimization returns
\[
\rho=14.11,
\qquad
\nu=-4.71.
\]
The corresponding MPC policy is then tested on the remaining 399 wind realizations. One of the trajectories corresponding to a wind realization still enters the low-connectivity zone, as shown in Panel (b) of Figure~\ref{fig:first_pick}. This violating realization is picked and inserted into $T$. 

The training set now contains two wind realizations. The learning problem is solved again, minimizing the average traveled distance over these two cases while enforcing avoidance of the low-connectivity zone for both. The new values are
\[
\rho=7.59,
\qquad
\nu=-4.55.
\]
With this policy, none of the trajectories corresponding to the remaining 398 wind realizations enters the low-connectivity zone, see Panel (c) of Figure~\ref{fig:first_pick}. 

The final training set has cardinality
\[
|T|=2.
\]
Using confidence parameter
\[
\beta=10^{-5},
\]
Theorem~\ref{thm:p2l} gives
\[
R(\pi(T)) \leq 4.8\%.
\]
Here, the risk is the probability that, under a fresh and previously unseen wind realization, the aircraft trajectory generated by the designed policy enters the low-connectivity zone. Thus, with confidence $1-10^{-5}$, the violation probability of the learned MPC policy is no larger than 4.8\%. 

\section{Post-Design Certification of Additional Properties}
\label{sec:post_design}

The P2L run described above was carried out with one specific property in mind, namely avoidance of the original low-connectivity zone. Once the MPC policy has been designed, however, one may want to certify other properties without changing the control policy. This is the role of the post-design evaluation step. The idea is the one formalized by P2L$^+$ in \cite{P2L2025ArXiv}: a property can be evaluated after the design by adding to the original compression set the scenarios, among those not already in the compression set, that violate the new property.

Let $\pi(T)=\MPC(\rho,\nu)$ be the policy returned by the P2L procedure and let $T$ be the final training set. In the numerical experiment of Section~4, $|T|=2$. Consider now a new property $P$, different from the property used during the design phase. For each wind realization $\delta_i$, let
\[
\psi_P(\pi(T),\delta_i)\in\{0,1\}
\]
be the corresponding indicator, where $\psi_P(\pi(T),\delta_i)=1$ means that the closed-loop trajectory generated by $\pi(T)$ under $\delta_i$ satisfies $P$. Define
\[
U_P=\{\delta_i\in D\setminus T: \psi_P(\pi(T),\delta_i)=0\}.
\]
Thus, $U_P$ contains the scenarios, not already in the compression set, for which the new property is violated. The empirical count $|U_P|$ contributes to the new complexity according to formula 
\[
k_P=|T|+|U_P|.
\]
The following result is the post-design specialization of P2L$^+$ and it will be used below in the current example. 

\begin{theorem}[Post-design P2L$^+$ certificate]
	\label{thm:post_design}
Let $(\pi(T),T)$ be the output returned by P2L when applied to a dataset $D$ of $N$ independent uncertainty realizations. Let $P$ be any additional property of interest and let $U_P$ be the set of elements in $D\setminus T$ that violate $P$ under policy $\pi(T)$. Define
	\[
	R_P(\pi(T))=\mathbb P_{\delta}\{\delta:\psi_P(\pi(T),\delta)=0\}.
	\]
	Then, for any confidence parameter $\beta\in(0,1)$, with confidence at least $1-\beta$ with respect to the random extraction of the dataset,
	\[
	R_P(\pi(T))\leq \varepsilon(|T|+|U_P|,\beta,N),
	\]
	where $\varepsilon(\cdot,\beta,N)$ is the same function as in Theorem~\ref{thm:p2l}.
\end{theorem}

The theorem gives a direct operational rule. To certify a new property after the MPC design, one simulates the closed loop on the scenarios in $D\setminus T$, counts how many of them violate the property, adds this number to $|T|$, and evaluates the same bound used earlier in the paper. The empirical violation count is therefore the starting point of the certificate, but the final statement concerns the probability of violation under a fresh wind realization. This procedure is illustrated for the flight problem in the next subsections. 

\subsection{A More Conservative Connectivity Requirement}

\begin{figure}[ht]
	\centering
	\includegraphics[width=10cm]{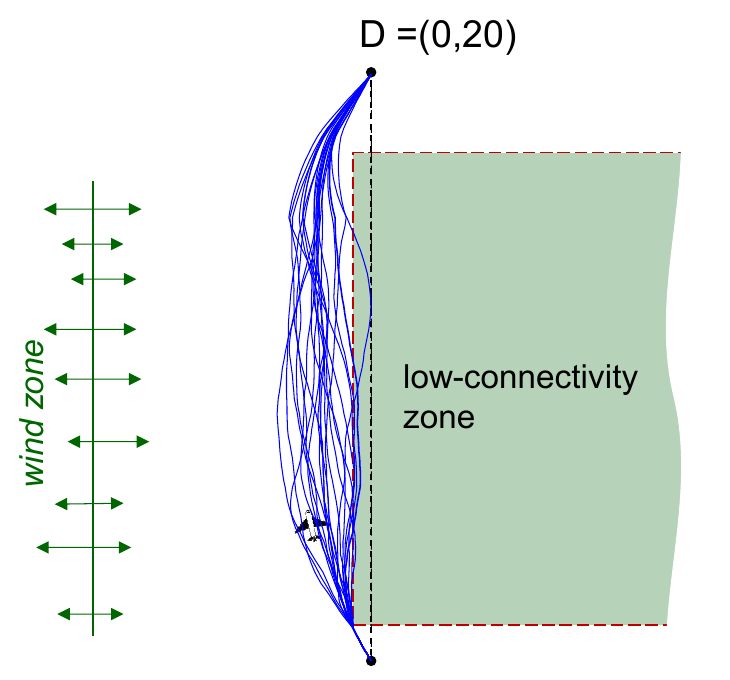}
	\caption{Post-design evaluation for an enlarged low-connectivity zone. The green region represents the more conservative connectivity constraint and corresponds to the original region shifted left by $0.125$. A representative subset of the trajectories generated by the final MPC policy is shown.}
	\label{fig:post_green_any}
\end{figure}

We first reconsider the connectivity constraint. The original design required the aircraft to avoid the low-connectivity zone represented in Figure~\ref{fig:geometry}. Suppose now that connectivity degrades gradually around this region and that, to remain on the safe side, the aircraft should avoid a larger set. This enlarged set is represented in green in Figure~\ref{fig:post_green_any}. The policy is not redesigned and we want to evaluate the probability of violating this more stringent constraint. 

The new property is
\[
P_{\mathrm{green}}: \quad \text{the trajectory does not enter the enlarged green region.}
\]
Among the scenarios in $D\setminus T$, 14 trajectories enter the enlarged region. Therefore
\[
k_{\mathrm{green}}=|T|+14=2+14=16.
\]
Using Theorem~\ref{thm:post_design} with $N=400$ and $\beta=10^{-5}$ gives
\[
R_{\mathrm{green}}(\pi(T))\leq 11.4\%.
\]
This is a bound on the probability, certified with confidence $1-10^{-5}$, that a new wind realization causes the trajectory of the learned MPC policy to enter the enlarged low-connectivity region.

Suppose now that losing connectivity for a few sampling instants does not cause a major operational problem. We therefore fix a tolerance threshold of 10 consecutive sampling instants and declare a trajectory violating only if it remains inside the enlarged green region for at least 10 consecutive instants. This gives the property
\[
\begin{array}{ll}
	P_{\mathrm{green},10}: &
	\text{the trajectory does not remain in the enlarged green region for 10 consecutive} \\
	& \text{sampling instants or more.}
\end{array}
\]
In this case, the post-design count gives 3 violating scenarios in $D\setminus T$, and hence
\[
k_{\mathrm{green},10}=|T|+3=2+3=5.
\]
With $N=400$ and $\beta=10^{-5}$, Theorem~\ref{thm:post_design} yields
\[
R_{\mathrm{green},10}(\pi(T))\leq 6.4\%.
\] 

\subsection{Certification of Traveled Distance}

The same post-design mechanism can be applied to performance properties. Let $L(\pi,\delta)$ denote the total distance traveled by the closed-loop trajectory generated by policy $\pi$ under wind realization $\delta$. For a threshold $\gamma$, define
\[
P_\gamma: \quad L(\pi,\delta)\leq \gamma.
\]
The risk associated with this property is
\[
R_\gamma(\pi(T))=\mathbb P_{\delta}\{\delta: L(\pi(T),\delta)>\gamma\}.
\]
Equivalently, if $F_L$ is the cumulative distribution function of the random traveled distance, then
\[
F_L(\gamma)=1-R_\gamma(\pi(T)).
\]

For example, consider the threshold
\[
\gamma=20.05, 
\]
which adds 0.05 to the minimal distance from Origin to Destination. Among the scenarios in $D\setminus T$, 105 trajectories travel a distance larger than 20.05. Thus,
\[
k_{20.05}=|T|+105=2+105=107.
\]
Using Theorem~\ref{thm:post_design} with $N=400$ and $\beta=10^{-7}$ gives
\[
R_{20.05}(\pi)\leq 41.2\%,
\]
and therefore
\[
F_L(20.05)\geq 58.8\%
\]
with confidence at least $1-10^{-7}$.

\begin{figure}[ht]
	\centering
	\includegraphics[width=10cm]{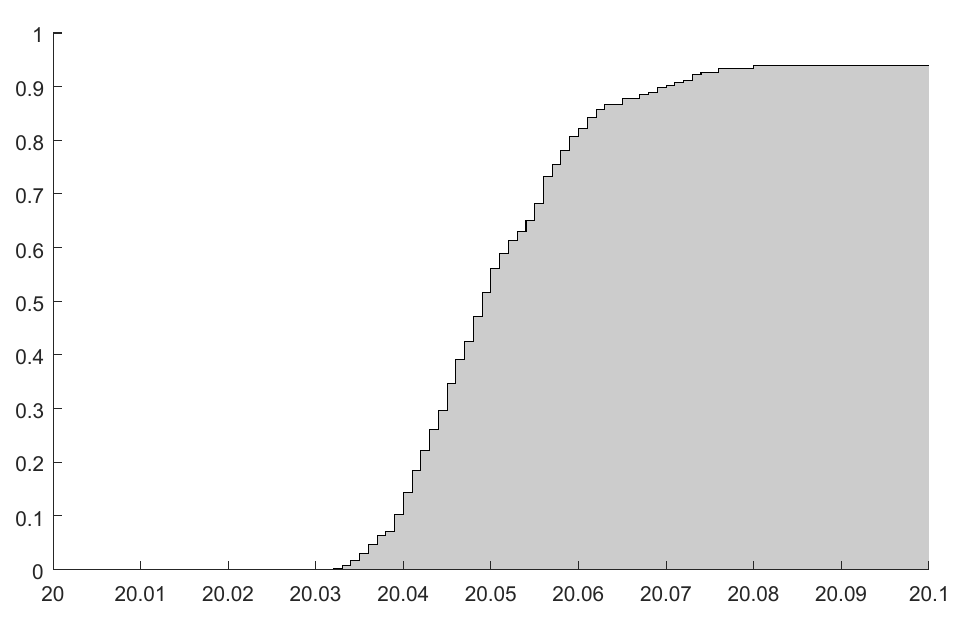}
	\caption{Certified lower bound on the cumulative distribution function of the traveled distance. The bound is obtained by applying the post-design certificate at 100 distance thresholds with confidence parameter $10^{-7}$ for each threshold, yielding an overall confidence level at least $1-10^{-5}$. %The figure also displays the true cumulative distribution function in blue.
	}
	\label{fig:post_cdf_bound}
\end{figure}

Repeating the same computation for a grid of thresholds $\gamma_1,\ldots,\gamma_{100}$ produces a lower bound on the cumulative distribution function of the random traveled distance. Specifically, for each threshold $\gamma_j$, let
\[
U_j=\{\delta_i\in D\setminus T: L(\pi,\delta_i)>\gamma_j\},
\qquad
k_j=|T|+|U_j|.
\]
Then
\[
F_L(\gamma_j)\geq 1-\varepsilon(k_j,10^{-7},400),
\qquad j=1,\ldots,100,
\]
holds simultaneously for all 100 thresholds with confidence at least $1-10^{-5}$, by the union bound. The resulting lower bound for the flight problem is shown in Figure~\ref{fig:post_cdf_bound}. 

\section{Discussion and Conclusions}

We have addressed in this paper a specific flight-control problem to illustrate how P2L can be used to calibrate an MPC policy and certify its closed-loop behavior. The flight-control application considered here is merely intended as an example of a general methodology. In this methodology, the finite-horizon optimization problem defines a family of MPC policies, while P2L operates at the policy level by selecting the parameters that achieve the desired closed-loop behavior over the available disturbance scenarios.

The same methodology can be used in other MPC settings where robustness against disturbances is required. A natural example is tube MPC. In tube MPC, one considers a tube around a nominal trajectory designed so that the deviation induced by the disturbance remains inside the tube. The tube is then required to lie within the constraint-admissible region, thereby ensuring constraint satisfaction for all disturbances in the prescribed uncertainty set. This is a worst-case construction. A similar approach can be pursued using the P2L methodology of this paper, if a relaxation of the worst-case requirement is tolerable. For instance, one may introduce a constraint enlargement, or equivalently a tightening of the nominal admissible region, as in tube MPC, when solving the FHOCP. However, rather than fixing this enlargement a priori from a worst-case disturbance model, one could calibrate it with P2L in the same spirit in which the parameters $\rho$ and $\nu$ have been calibrated here. This would provide a probabilistic, data-driven alternative to worst-case tube design, while preserving the possibility of certifying the resulting closed-loop behavior. Along this approach, another important advantage is that the disturbance model used to generate the scenarios can be arbitrarily complex, and no explicit knowledge of its probability distribution is required beyond the availability of independent draws. Thus, the proposed methodology can be applied with disturbance scenarios generated by detailed simulators, high-fidelity models, or other complex procedures, while still providing rigorous and typically tight certificates for the resulting closed-loop behavior.

Finally, the disturbance scenarios used by P2L need not be generated from a model. They may instead be obtained directly as observations. In this case, one can avoid specifying a probabilistic model for the disturbance and still obtain rigorous guarantees, thanks to the distribution-free nature of the P2L certificate. Moreover, since P2L does not require setting aside data for testing or calibration, all available observations can be used in the construction and certification of the policy.

\bibliographystyle{unsrt}
\bibliography{biblio}

\end{document}